\documentclass[prd,nofootinbib,preprintnumbers,floatfix]{revtex4}  
\usepackage[plainpages=false,  colorlinks=true, anchorcolor=blue, linkcolor=blue, citecolor=blue, bookmarks=false]{hyperref}
\usepackage{natbib}
\usepackage{graphicx}
\newcommand{\rthis}[1]{\textcolor{black}{#1}}
\begin{document}

\title{Combined significance of spatial coincidence of  high energy neutrinos from PSR B1509-58 by Super-Kamiokande and MACRO }

\author{Shantanu \surname{Desai}}%
\altaffiliation{E-mail: shntn05@gmail.com}

\date{\today}
\affiliation{Department of Physics, Indian Institute of Technology, Hyderabad, Kandi, Telangana-502285, India}


\begin{abstract}

In their searches for astrophysical point sources of high energy neutrinos, both the  Super-Kamiokande and MACRO neutrino detectors saw the largest angular excess  from the same source, viz. PSR B1509-58. We estimate the  probability   for the   observed number of events by {\it both} Super-Kamiokande and MACRO  to be  a chance coincidence  due to  atmospheric neutrino background. We find  that this probability is about 0.4\%, corresponding to 2.6$\sigma$ significance. We also propose some additional tests  to ascertain if this excess corresponds to an astrophysical signal or is only a background event.

\end{abstract}

\maketitle

\section{Introduction} \label{sec:intro}

There are plenty of examples in electromagnetic astronomy, when one could harness a lot more science from astronomical observations of a given source by combining the data from multiple telescopes, instead of using the data from only a single telescope. A non-exhaustive list  which demonstrates the power of combining data from multiple telescopes throughout the electromagnetic spectrum  includes VLBI~\citep{Ryle}, optical interferometry~\citep{Monnier}, black hole image in M87 with the  Event Horizon Telescope~\citep{EHT}, TeV gamma-ray astronomy with upcoming detectors such as CTA~\citep{CTA}, etc. The main reason for building multiple ground-based  interferometric gravitational wave  detectors around the globe was to enhance the detection significance and  and get better pointing accuracy, as compared to using data from only a single detector~\citep{LIGO1,LIGO2}. The search for nanoHz gravitational waves is also being done by pooling the data from multiple radio telescopes~\citep{Hobbs}. Recently,~\citet{Oyama} has demonstrated the power of combining data from multiple neutrino  detectors, and has shown that the detection significance of  upward muons from SN1987A using both Kamiokande and IMB gets enhanced (2.8$\sigma$), as compared to analyzing data from each detector separately. 

In this work, we try to combine the data from two other neutrino detectors, namely MACRO and Super-Kamiokande (Super-K, hereafter) to jointly assess the detection significance of PSR B1509-58 using both the detectors.  Super-K  has carried out a number of searches for astrophysical neutrinos in the GeV to TeV regime in the first phase of the experiment from April 1996 to July 2001 using upward going muons~\citep{superknuastro,superkgrb,superkwimp,superkshowering,Swanson},  to supplement the other path-breaking results on atmospheric and solar neutrinos~\citep{Kajita}. Around the same time, the MACRO detector has also carried out searches for astrophysical point sources of neutrinos in the GeV-TeV energy range  using data from 1989 to 2000 ~\citep{MACROnuastro,Montaruli}. The upward-going  muons for both the detectors are produced by high energy neutrinos undergoing charged current interactions in the rock surrounding the detectors. Both Super-K and MACRO have similar muon energy threshold ($~\sim$ 1 GeV), angular resolution (~$\sim 1^{\circ}$),  effective area ($\sim$ 1000 $m^2$),  nearly the same latitude in the northern hemisphere ($36^{\circ}$ for Super-K versus  $42^{\circ}$ for MACRO). Therefore, they have a similar sensitivity to astrophysical neutrinos in the GeV-TeV energy range to sources in the same part of the southern sky. Both the detectors also have  four years of overlap  from July 1996 to  December 2000. 

Although, neither Super-Kamiokande nor MACRO have detected a statistically significant excess of upward muons from any  astrophysical source, both the detectors saw the largest excess from the same astrophysical source in a cone of half-angle equal to $3^{\circ}$, viz. PSR B1509-58~\citep{superknuastro,Montaruli} (also T. Montaruli, private communication in 2006). The source PSR B1509-58 is a pulsar  with  RA=15:13:55.8 (hr:mt:sec) and DEC=$-59.14^{\circ}$, and located at a distance of 4.4 kpc (cf. v1.66 of ATNF pulsar catalog~\citep{ATNF}).
In this work we discuss if the detection significance gets enhanced, after using data  from both these detectors. For our analysis, we use the same methodology as in ~\citet{Oyama}.

This manuscript is structured as follows. Our analysis and results are discussed in Section~\ref{sec:analysis}. A discussion of our results can be found in Sect.~\ref{sec:discussions}. We conclude in Sec.~\ref{sec:conclusions}.

\section{Analysis}
\label{sec:analysis}

\subsection{Super-K results}
Super-K carried out a search for astrophysical point sources of neutrinos from 52 distinct sources  using 2359 upward muon events detected in the first phase of the experiment from April 1996-July 2001,  with a muon  energy threshold of 1.6 GeV~\citep{superknuastro}. The upward muon sample consists of 1892 through-going and 467 stopping muons. \rthis{Out of 1892 through-muons, 318 are showering  through-going muon events and the rest are  non-showering events~\cite{superkshowering}.}
No statistically significant excess was seen from any of the sources. However, the largest excess was seen for PSR B1509-58 with a total of 6 signal events in a cone of half-angle $3^{\circ}$ with the expected atmospheric neutrino background of 2.8 events. This cone contains 90\% of the signal for an $E^{-2}$ neutrino spectrum~\citep{Montaruli}.
The Poisson probability of getting a number greater than or equal to the number of observed events, considering this background is equal to $1- \sum \limits_{i=0}^5$ Pois ($i | 2.8$), where Pois ($i | 2.8$) is the Poisson probability mass function with background of 2.8 and mean equal to $i$ (cf. Table~\ref{tab1} for the full expression). The aforementioned Poisson probability is equal to 0.065. Accounting for the fact that Super-K looked at 52 sources, and 
thereby multiplying 0.065 by 52 to take into the trials factors, one finds that the fluctuation is consistent with background. Super-K also did a search using the upward showering muon subset from the same source~\citep{superkshowering}. In this case, the observed showering muons within the same cone is equal to 1, corresponding to a background of 0.4 events~\citep{superkshowering}, \rthis{indicating that that there is no significant excess}.

\subsection{MACRO results}
The MACRO results for searches from astrophysics starting from March 1989 to September 1999 and December 2000 have been presented in ~\cite{MACROnuastro} and ~\cite{Montaruli}, respectively. The final search has been  done 
with 1388 upward muons. The largest excess in the $3^{\circ}$ half-angle cone was again seen for PSR B1509-58 with a total of 10 events compared to expected atmospheric background of 2 events.  The Poisson probability of getting a number greater than or equal to number of observed events given this background is equal to $1- \sum \limits_{i=0}^9$ Pois ($i | 2$), which comes out to  $4.6 \times 10^{-5}$. Accounting for the trials factors for this is slightly tricky since the total number of neutrino sources  which have been looked at for the full analysis until December 2000 is not specified~\cite{Montaruli}. A total of 42 astrophysical point sources were searched for the analysis until Sept 1999~\cite{MACROnuastro}. The final analysis argues that to take into account the  trials factor, one needs to consider the event directions around all the 1388 detected upward muons~\cite{Montaruli}.   We multiply the aforementioned probability by the same trials factor of 1388, giving rise to the final probability of it being a chance fluctuation  equal to 0.064. This corresponds to a Z-score  statistical significance of only 1.5$\sigma$, following the prescription in ~\cite{Cowan}, which is not significant.

\subsection{Results from combined analysis}
The Super-K and MACRO results for searches from PSR B1509-32 are summarized in Table~\ref{tab1}. We now combine the detection probabilities from both the detectors, to estimate the joint significance. Since the MACRO and Super-K searches were carried out independently, we can assume that the combined probability is product of individual probabilities, since the results are independent of each other. We use the same trials factor of 1388 as in the latest MACRO result~\citep{Montaruli}.
As Super-K and MACRO are sensitive to nearly the same part of the southern skies, we assume that the trials factor of 1388 used by MACRO encompasses all the sources searched for by Super-K.
Therefore, the final probability that both the Super-K and the MACRO neutrino events in the direction of PSR B1509-58 are a chance fluctuation is given by the product of the individual probabilities and the trials factors used:
$$ P = 0.065 \times 4.6 \times 10^{-5} \times 1388 = 4.1 \times 10^{-3}$$
Therefore, the joint detection significance probability is equal to  0.4\%, corresponding to a Z-score significance of  2.6$\sigma$ using the prescription in ~\cite{Cowan}.

\begin{table*}[h]
\caption{\label{tab1}
Number of upward-going muon events detected by Super-K (in the first phase from 1996-2001) and MACRO (using data from 1989-2000) from PSR B1509-54. The first column indicates the neutrino detector used. The second column shows the observed upward muon events in a $3^{\circ}$ cone half-angle. The third column indicates
the atmospheric neutrino background with the same angular region. The calculation and 
the Poisson probability $N \geq N_{cand}$ are shown in the last two columns. The Poisson distribution function  Pois($N | \mu$) is given by $\frac{\mu^N e^{-\mu}}{N!}$.  
}
\begin{center}
\begin{tabular}{lllll}
\hline
\hline
 Detector & Signal &  Background & Calculation &  Prob ($N \geq N_{sig}$) \\
\hline
Super-K &~~~~~6&~~~~~~~2.8&~~~$1- \sum \limits_{i=0}^5$ Pois ($i|$2.8)&~~~0.065\\ \\
MACRO  &~~~~~10&~~~~~~~2&~~~ $1- \sum \limits_{i=0}^9$ Pois ($i|$2)&~~~ $4.6 \times 10^{-5}$ \\

\hline
\hline
\end{tabular}
\end{center}
\end{table*}

\subsection{Possible systematic errors}
\rthis{Therefore prima facie, the combined statistical significance  after combining the upward muon data from MACRO and Super-K in the direction of PSR B1509-58 is 2.6$\sigma$. Here, we discuss possible systematics that are not accounted for in our analysis, but which could degrade the statistical significance. Super-K detects three kinds of upward muons. Although, the angular resolution of all the three muons is about the same, viz.  $1.4^{\circ}$ (showering), $1.3^{\circ}$ (non-showering thrugoing), $2.4^{\circ}$ (stopping)~\cite{superkshowering},
the mean angle between the parent neutrino and the reconstructed upward muons changes a lot across  the samples. For upward stopping muons, the mean separation for an atmospheric neutrino spectrum is about $8.7^{\circ}$, whereas it is about $2.1^{\circ}$ and $2.9^{\circ}$ for non-showering and showering thrugoing muons, respectively. Therefore, depending on how many  of the six  upward muons seen in Super-K are stopping muons, the statistical significance could get degraded and one may need to consider  larger cone angles.  At this moment the only  information which is available (from published Super-K data) is that out of the six observed events, one event is an upward showering muon~\cite{superkshowering}. }

\rthis{Another possible systematic for the Super-K sample is that the all the three kinds of upward muons with $\cos(\theta)>-0.1$, ( where $\theta$ is the upward muon zenith angle) are contaminated by downward going muons from the horizon due to multiple-coulomb scattering and finite angular resolution of the fitters~\cite{superkshowering}. Although, it is straightforward  to statistically estimate this background and subtract it for oscillation analysis~\cite{superkshowering}, one cannot cull such near-horizon muons on an event by event basis during  astrophysical searches. For astrophysical searches using Super-K-1 data, a weighting factor was applied for upward muons with $\cos (\theta) > -0.1$ in each of the myriad astrophysical searches carried out~\cite{superkgrb,superknuastro,superkwimp,superkshowering}.
If any of the six Super-K events have $\cos(\theta)>-0.1$, that would degrade the statistical significance of our result. Again, the zenith angle distribution of these six Super-K events are not publicly available.}

\rthis{For selecting the  MACRO upward muon sample used for astrophysical searches, no cut was imposed on the minimum amount of material a muon has to pass through, unlike in the oscillation analysis~\cite{MACROnuastro}. Therefore, there could be some residual contamination from photo-produced pions in this sample~\cite{MACRObkgd}.  Furthermore, as discussed in ~\cite{MACROnuastro}, the MACRO dataset also contains events having an interaction vertex in the lower half of the detectors, thereby increasing the number of possible background events. Similar to Super-K, based on the publicly available information it is not possible to assess if any of the MACRO upward muons seen in the direction of the pulsar are the aforementioned background events. If it could be ascertained that any of the 10 events are not neutrino-induced, that would again degrade the joint statistical significance estimated in this work.}

\section{Discussions}
\label{sec:discussions}
Therefore, we find that the probability that both Super-K and MACRO would detect the observed events as a fluctuation of background is about 0.41\%, corresponding to a significance of 2.6$\sigma$. Although this does not pass the 5$\sigma$ discovery criterion in high energy physics~\cite{Lyons},  we propose some additional tests which, could be carried out to ascertain if  both Super-K and MACRO have detected astrophysical neutrinos from PSR B1509-58.

There are two types of astrophysical point sources: steady-state and transient sources. Pulsars have been known to be promising sources of high energy neutrinos for  a long time~\citep{Helfand79}. We also note that theoretical models for steady-state neutrino emission from PSR B1509-58 have been proposed, \rthis{in case protons could be accelerated to 1 PeV~\citep{Link}}, although the event rates in MACRO/Super-K size detectors are negligible, since the models in ~\cite{Link} obtain an event rate of only $\sim$ 5 events/km$^2$/year.
However, if PSR B1509-58 was a steady state source of neutrinos, then one would have  expected that the statistical significance would accumulate in Super-K with increased exposure. Super-K is still taking taking data. Although there has been no follow-up result from this pulsar with Super-K,  subsequent searches for astrophysical neutrinos from Super-K after July 2001 have not found any statistical significant excess from  any point source~\citep{Thrane09,sknuastro17}. Other upward muon detectors in the northern hemisphere such as ANTARES or Baikal have also not seen any statistical significant excess from the direction of this pulsar~\citep{Antares,Baikal}. Although the IceCube detector is located in the Southern Hemisphere, it is sensitive to TeV neutrinos from the direction of PSR B1509-52 and no statistically significant excess has been reported~\citep{IceCube}.

The other possibility is that this pulsar is a transient source of astrophysical neutrinos. We note that a  TeV gamma ray flare was detected (at 4.1$\sigma$) from this pulsar in 1997 from the Cangaroo Atmospheric Cherenkov telescope. \rthis{Therefore, in case this TeV flare was produced from any hadronic acceleration mechanism, there would also be associated neutrinos from pion decay, observed  in coincidence with the flare. Therefore, one could carry out temporal searches using both MACRO and Super-K detectors around this TeV flare.}
However, unlike ~\citet{Oyama} a temporal search cannot be done with publicly available 
data, since both the Super-K and  MACRO upward muon dataset is proprietary and the MJD (or the observed RA, DEC)  of the upward muon dataset is not publicly available.
Such a search can only be done by both the collaborations. Another possibility would be to carry a joint search with the Baksan neutrino telescope, which is located at approximately the same latitude as Super-K and MACRO, and is sensitive to upward muons with similar energy threshold as in Super-K/MACRO. Baksan has been taking data since 1977~\citep{Baksan99} and is still operating today~\citep{Baksan21}.  It would be very interesting if the Baksan detectors also saw an excess from this source.\footnote{We could not find any publication on all sky searches for high energy astrophysical point source of neutrinos from Baksan, but  to the best of my knowledge, it was  taking data between 1996 and 2000.} \rthis{Nevertheless, it is important to carry out a temporal search to supplement the current analysis, independent  of any electromagnetic transient flares observed from this pulsar or any prejudice from theoretical models.}

\section{Conclusions}
\label{sec:conclusions}
Super-K and MACRO  have carried out searches for point sources of astrophysical neutrinos with upward muons  produced from high energy neutrinos in the GeV-TeV energy range.
Although no astrophysical source was detected from both the detectors, one intriguing  observation was that both the detectors saw the largest excess (from all-sky searches within a $3^{\circ}$ angular region) from the same astrophysical source, viz. PSR B1509-58.  \rthis{However, the significance of this excess is negligible when each dataset is analyzed separately.}
We try to assess the combined probability that the upward muons detected by both  MACRO and Super-K within the above angular region around PSR B1509-58 are a fluctuation of the background, using publicly available data. We follow the same prescription as in ~\citet{Oyama}, who combined data from the IMB and Kamiokande detectors to assess the joint significance of detection of upward muons from SN 1987A.

We estimate the chance coincidence of events seen by Super-K and MACRO to be a fluctuation from atmospheric neutrino background to be 0.41\%. This corresponds to a Z-score significance of 2.6$\sigma$  \rthis{and therefore the detection significance  gets enhanced, when one combines the data from both Super-K and MACRO. 
We have also discussed potential sources of systematic errors in both the Super-K and MACRO upward muon dataset, which could
potentially degrade the aforementioned significance.}

We have also argued why this source cannot be a steady state source of high energy astrophysical neutrinos, since  this signal would have seen by other detectors (including Super-K) with additional exposure. In order to ascertain if PSR B1509-58 was a transient source of astrophysical neutrinos,  one needs to extend this analysis by carrying out a temporal search using the observed arrival time of MACRO and Super-K upward muon events.  Another possibility would be  to include data from the Baksan neutrino detector, which was also taking data during the overlapping period between MACRO and Super-K.  

\rthis{Many new detectors of GeV neutrinos such as DUNE~\cite{DUNE}, INO~\cite{INO}, JUNO~\cite{JUNO}, PINGU~\cite{PINGU}, Hyper-K~\cite{Hyper-K} etc  are soon going to come online and start taking data. The history of electromagnetic and gravitational wave astronomy has shown   that the combining data from multiple  detectors has helped us learn a lot about the Physics of the sources, as compared to each detector separately. We hope this lesson can be extended to neutrino astrophysics in the coming decade, by combining data from the different neutrino detectors.}

\section*{Acknowledgments}
I am grateful to Teresa Montaruli for pointing out to me (about 15 years ago)  that Super-K and MACRO have seen the largest excess for the same source, viz PSR B1509-58, and to John Beacom for ensuing discussions on this. I am thankful to all the colleagues (in particular Alec Habig) of the upward muon  group  in  Super-Kamiokande for many years of fruitful discussions. \rthis{We also thank the anonymous referees for useful constructive feedback on this work}.

\bibliography{sample631}
%\facilities{Super-Kamiokande, MACRO}
%\software{scipy}
\end{document}